\documentclass[aps,prl,superscriptaddress,twocolumn,notitlepage,nofootinbib,longbibliography]{revtex4-2}

\usepackage{lineno}
\usepackage[dvips]{graphicx}
\usepackage{braket}
\usepackage{dcolumn}
\usepackage{comment}
\usepackage{bm}
\usepackage{amsmath,amsthm,amssymb}
\usepackage{epstopdf}
\usepackage{mathrsfs}
\usepackage{color}
\usepackage{xcolor}
\usepackage{bbold}
\usepackage{mathtools}
\usepackage[sort&compress]{natbib}
\usepackage{hyperref}
\usepackage{cleveref}
\hypersetup{colorlinks=true,linkcolor=blue,citecolor=blue,urlcolor=black}

\definecolor{azure}{rgb}{0.0, 0.5, 1.0}
\definecolor{amber}{rgb}{1.0, 0.49, 0.0}
\definecolor{red}{rgb}{1.0, 0.1, 0.0}
\definecolor{forestgr}{rgb}{0.13, 0.55, 0.13}

\begin{document}

\title{Self-similarity in pandemic spread and fractal containment policies}

\author{Alexander F.~Siegenfeld}

\affiliation{Department of Physics, Massachusetts Institute of Technology, Cambridge, MA}
\affiliation{New England Complex Systems Institute, Cambridge, MA}
\author{Asier Pi\~neiro Orioli}
\affiliation{New England Complex Systems Institute, Cambridge, MA}
\author{Robin Na}
\affiliation{Sloan School of Management, Massachusetts Institute of Technology, Cambridge, MA}
\author{Blake Elias}
\author{Yaneer Bar-Yam}
\affiliation{New England Complex Systems Institute, Cambridge, MA}

\date{\today}

\maketitle

\textbf{Although pandemics are often studied as if populations are well-mixed, disease transmission networks exhibit a multi-scale structure stretching from the individual all the way up to the entire globe~\cite{eubank2004modelling,Watts2005,colizza2007reaction,Gosme2009,goldstein2009reproductive,ajelli2010comparing,ball2015seven,Murillo2013,Yuan2023}. The COVID-19 pandemic has led to an intense debate about whether interventions should prioritize public health or the economy, leading to a surge of studies analyzing the health and economic costs of various response strategies~\cite{Karatayev2020,Acemoglu2021,Alvarez2021,brodeur2021literature,yaesoubi2021adaptive,kompas2021health,Neary2022}.  
Here we show that describing disease transmission in a self-similar (fractal) manner across multiple geographic scales allows for the design of multi-scale containment measures that substantially reduce both these costs.  We characterize response strategies using \textit{multi-scale reproduction numbers}---a generalization of the basic reproduction number $R_0$---that describe pandemic spread at multiple levels of scale and provide robust upper bounds on disease transmission.  Stable elimination is guaranteed if there exists a scale such that the reproduction number among regions of that scale is less than $1$, even if the basic reproduction number $R_0$ is greater than $1$.  We support our theoretical results using simulations of a heterogeneous SIS model for disease spread in the United States constructed using county-level commuting, air travel, and population data.}

\begin{figure*}
\centering
\includegraphics[width=\textwidth]{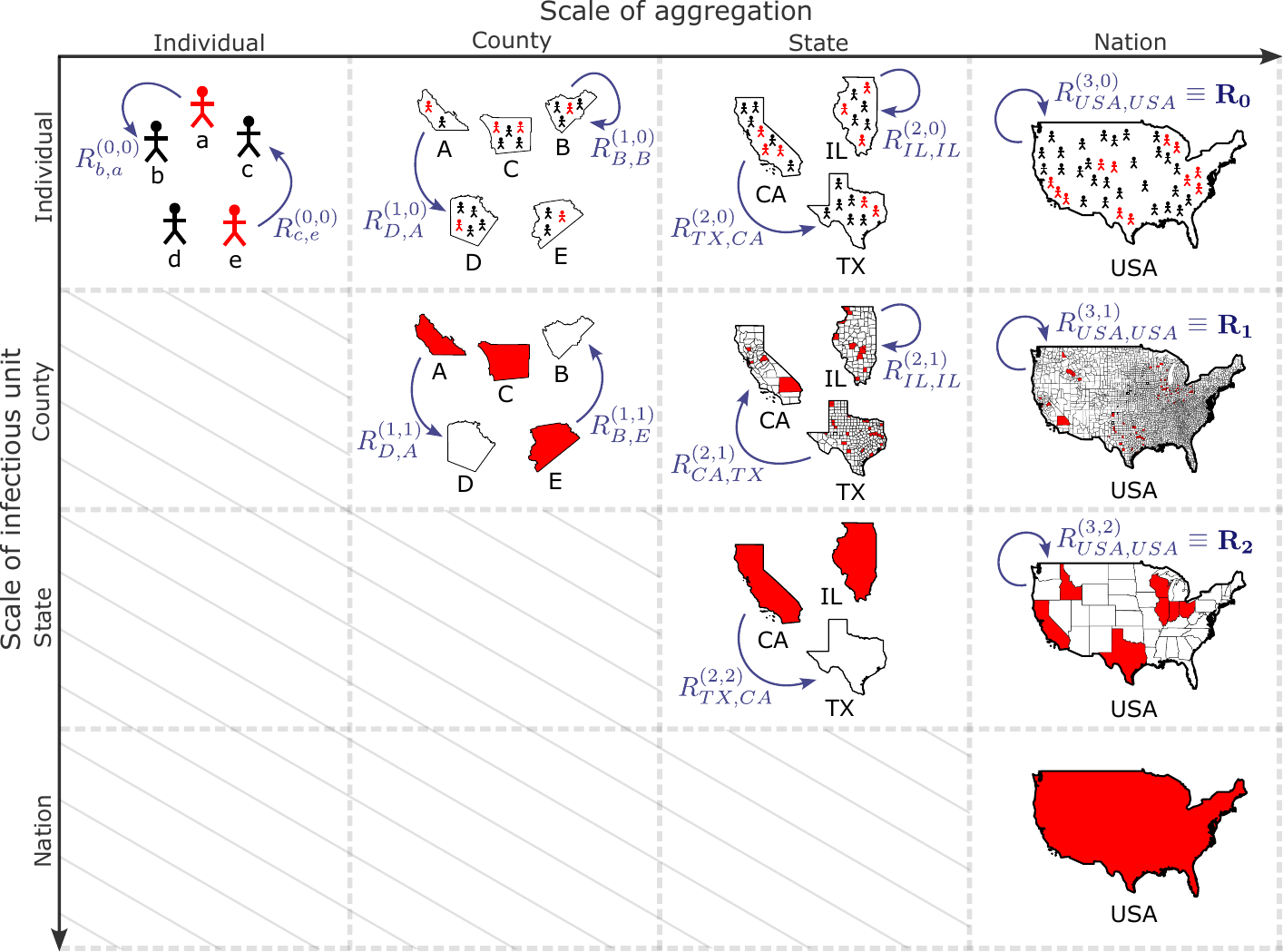}
\caption{\textbf{Disease spread at different levels of representation.} The figure illustrates infection processes as described by a next-generation matrix $R^{(m,n)}$. Here, $n$ is the scale of the infectious unit (vertical axis), and $m$ is the scale of aggregation (horizontal axis). The matrix elements $R^{(m,n)}_{ij}$ are dimensionless and describe transmission from $m$-level entity $j$ to $i$. For example, $R^{(0,0)}_{b,a}$ (top left corner) describes the probability that person $a$ will infect person $b$, whereas $R^{(2,1)}_{\text{CA,TX}}$ (second row, third column) describes the average number of counties in California that will be infected by an infected county in Texas before it recovers. Red regions or people represent infected entities, and arrows point in the direction of infection.} 
\label{fig:Rmn}
\end{figure*}

The dynamics of infectious disease spread is often encapsulated by the basic reproduction number, $R_0$, which determines whether an epidemic will grow, stabilize, or decline. If $R_0<1$, the number of infections will decline to the stable fixed point of zero cases, i.e. elimination.  We provide a mathematical and conceptual framework that generalizes the reproduction number to multiple scales.  The essential idea is shown in \cref{fig:Rmn}: disease transmission can be described between individuals (first row), in which transmission probabilities between individuals can be spatially aggregated into a nationwide basic reproduction number $R_0$ (last column). However, it can equally well be described between larger units such as counties (second row) and states (third row), in which transmission probabilities can be aggregated to county-to-county and state-to-state reproduction numbers ($R_1$ and $R_2$ in the last column of \cref{fig:Rmn}), respectively.  These coarse-grained descriptions of disease-spread between regions need not explicitly model the individuals within regions, in much the same way that epidemiological models that describe the spread of the disease at the level of individuals (e.g. labeling them as infectious, susceptible, recovered, etc.) are coarse-grained descriptions that ignore the fast and complicated spatio-temporal dynamics of the viral particles within each individual~\cite{nowak2000virus, perelson2002modelling, haseltine2008implications, mideo2008linking, Murillo2013, hernandez2020host}.   

$R_1$ and $R_2$ are similar to what is known in metapopulation models known as a global invasion threshold~\cite{colizza2007invasion,colizza2008epidemic}, as well as to reproduction numbers corresponding to smaller-scale social structures such as households~\cite{ball1997,fraser2007estimating,pellis2012reproduction}: if they are below $1$, then the disease will not be able to spread throughout the population and the disease-free state (elimination) will therefore be stable.  

Our analysis differs in that $R_1$ and $R_2$ capture the effects of spatiotemporally varying policies, i.e. different regions reacting differently at different times. Thus, rather than assuming that disease spread is limited only by the availability of susceptible hosts, we instead consider the case in which infected regions can ``recover'': i.e. if infected, a region chooses to temporarily implement measures until community transmission within the region has been fully contained, in much the same way that an immune system contains the spread of viral particles within an individual.  Which measures are appropriate will depend on the nature of the disease, and can include, for instance, improving indoor air quality, masks/respirators, testing/contact-tracing/quarantine, social distancing, travel reductions/testing at borders, and vaccination.  Due to strength of measures varying in time and space depending on which regions are infected, the system cannot be well-characterized by a single value of $R_0$.  Rather, the region-to-region reproduction number (e.g.~$R_1$ for a county-level policy or $R_2$ for a state-level policy) provides the appropriate criterion for the stability of elimination. 

\begin{figure*}
    \centering
    \includegraphics[width=\textwidth]{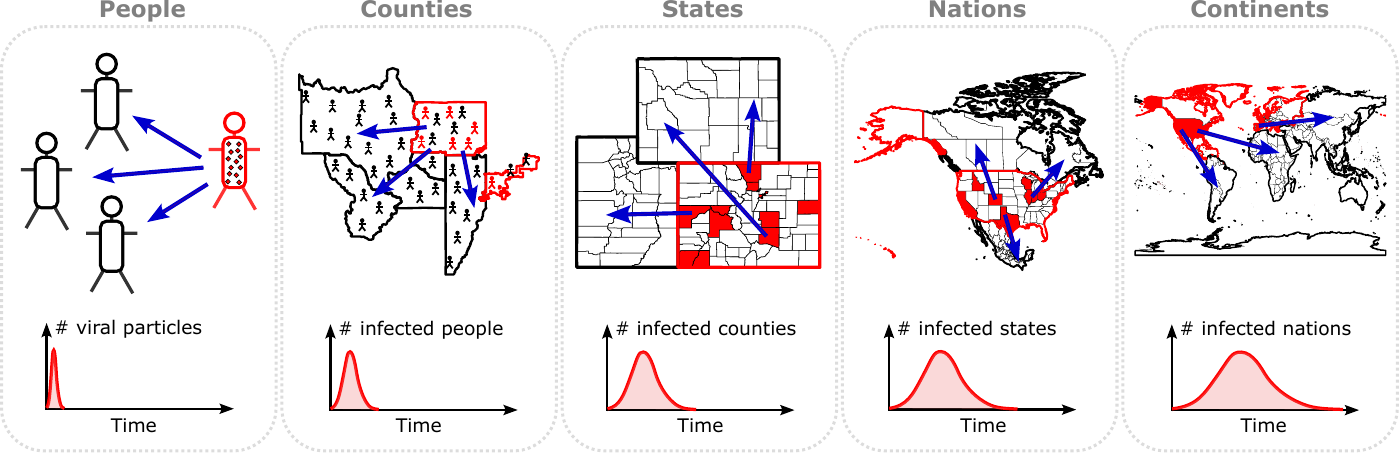}
    \caption{\textbf{Self-similar multi-scale approach.} During the rise and fall of a person's viral load over the course of an infection, that person will potentially infect other people.  Over longer time-scales, the same process can occur for counties, states, nations, and continents.
    In this figure, the scale of the infectious unit in each panel is equal to the aggregation scale of the previous panel (cf.~fig.~\ref{fig:Rmn}).}
    \label{fig:self-similarity}
\end{figure*}

This process of capturing smaller-scale behaviors within the parameters of a larger-scale model can be repeated in a self-similar way up to any desired scale.  As shown in \cref{fig:self-similarity}, viral particles are to individuals as infected individuals are to counties as infected counties are to states as infected states are to nations, etc.

This article is organized as follows.  We first introduce the self-similar multi-scale framework and give examples of policies that can be characterized by the multi-scale reproduction numbers $R_n$.  
We then show that these policies can achieve stable elimination, even when the pandemic has not been defeated worldwide and importations lead to new outbreaks with some frequency.  Finally, we discuss how these multi-scale policies can be optimized in a sequential/compartmentalized way.  We prove that two-scale policies are superior to single-scale/uniform policies and that costs can be further reduced by optimizing over three or more scales.   

We illustrate our theoretical conclusions by constructing an SIS metapopulation model among U.S. counties based on mobility and population data, with transmission probabilities that change over time as regions adapt their policies to the spatial distribution of infections.
The simulations demonstrate that even the simple multi-scale policies considered here can achieve elimination \emph{and} reduce costs for a range of potential pandemics evolving under approximately-realistic conditions.

\section{Multi-scale Framework}

Consider a population of individuals and a nested hierarchy of partitions of this population.
Throughout this work we will use the U.S. as a working example and consider four levels: individuals ($n=0$), counties ($n=1$), states ($n=2$), and the whole nation ($n=3$).  We note that not all levels need to be used, and that more levels (e.g.~households, neighborhoods, cities, etc.) could be added, but for ease of exposition we stick to one concrete example.  Note the nested structure: the U.S. consists of states, each of which consists of counties, each of which contains individuals.  
If we wish to model each nation separately, then $n=3$ would be the highest level.  However, we could also consider a collection of nations (\cref{fig:self-similarity}), in which case there could be additional levels $n=4$ corresponding to clusters of nearby nations (e.g.~nations on the same continent)  and $n=5$ corresponding to all of the nations under consideration (which could be the entire world, but could also be some subset of nations that wish to work together even if other nations are not cooperating).  

\begin{figure}
    \centering
    \includegraphics[width=\columnwidth]{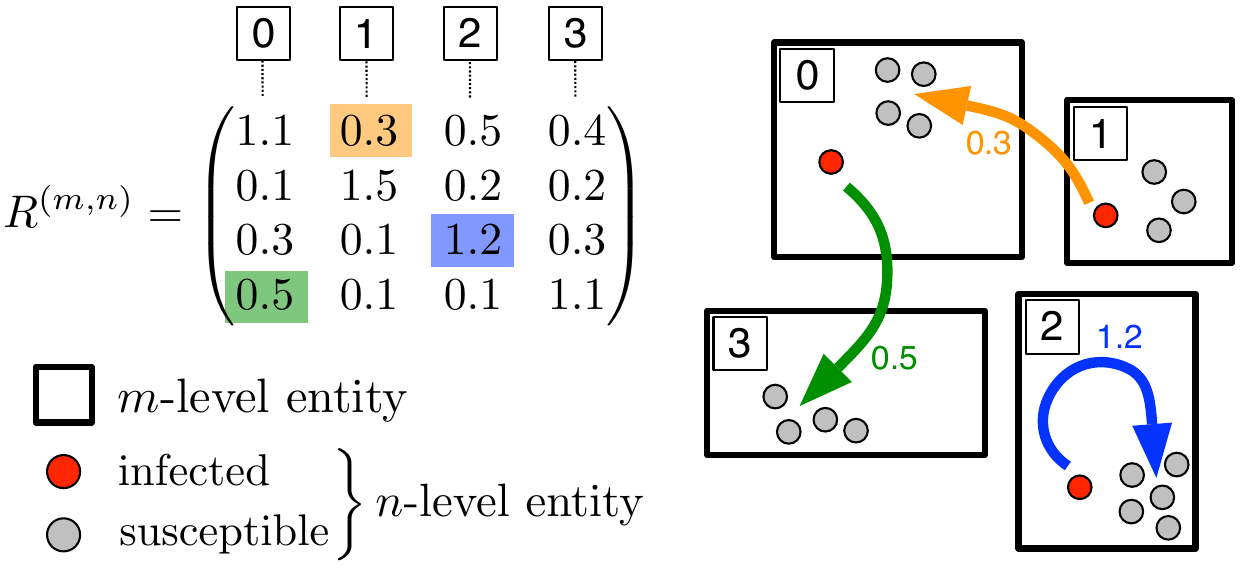}
    \caption{Pictorial depiction of $R^{(m,n)}_{ij}$ matrix.  
    }
    \label{fig:R}
\end{figure}

Suppose we wish to model a disease at the level of individuals (first row of \cref{fig:Rmn}).   We let $R^{(n,0)}_{ij}$ represent the average number of individuals that an infectious individual in region $j$ will infect in region $i$ (see \cref{fig:R}), where $n\geq 0$ denotes the scale of the regions.  The average includes the total number of infections caused by the individual in region $j$ from the time of their initial infection until their recovery. In other words, $R^{(n,0)}_{ij}$ denote next-generation matrices~\cite{diekmann2010construction}.  In our running example in which the highest scale is that of a single nation (the U.S.), the matrix $R^{(3,0)}$ would have only a single element, which would be the overall reproduction number $R_0$, while the next-generation matrix $R^{(1,0)}_{ij}$ would correspond to a metapopulation model~\cite{hanski1999metapopulation,Watts2005,colizza2007reaction,ball2015seven} denoting the spread of disease within and between U.S. counties (the diagonal elements representing the within-county reproduction numbers and the off-diagonal elements representing the inter-county reproduction numbers).  Likewise, the next-generation matrix $R^{(2,0)}_{ij}$ would describe a metapopulation model denoting the spread of disease within and between U.S. states.  $R^{(0,0)}_{ij}$ represents the expected number of times individual $j$, if infected, would infect individual $i$.   
We see then that the matrices $R^{(0,0)}$, $R^{(1,0)}$, $R^{(2,0)}$, and $R^{(3,0)}$ all describe the same system but at different levels of detail.  The finer-scale matrices can be coarse-grained to the larger-scale ones
\begin{equation}
\label{eq:spatial0}
R^{(0,0)} \rightarrow R^{(1,0)}\rightarrow R^{(2,0)}\rightarrow R^{(3,0)}
\end{equation}
such that the reproduction number $R_0$ is preserved. More precisely, all of the matrices in \cref{eq:spatial0} have the same largest eigenvalue, which is the appropriate definition of $R_0$ for heterogeneous disease spread~\cite{diekmann1990definition}; see Methods for more details.

We can also describe the disease at the level of counties (second row of \cref{fig:Rmn}).  We let $R^{(n,1)}_{ij}$ represent the average number of \textit{counties} that an infected \textit{county} in region $j$ will infect in region $i$ (\cref{fig:R}), where $n\geq 1$ denotes the scale of the regions.  In other words, $R^{(n,1)}_{ij}$ denote next-generation matrices but where counties rather than individuals are the infectious units.  (We define a county as infected if it contains at least one infected individual.)  In our running example, the matrix $R^{(3,1)}$ would contain a single element denoting the \textit{county-to-county reproduction number} $R_1$ of the U.S., while $R^{(2,1)}_{ij}$ would represent a metapopulation model describing the average number of counties in state $i$ expected to be infected by an infected county in state $j$.  $R^{(1,1)}_{ij}$ (analogous to $R^{(0,0)}_{ij}$) represents the expected number of times county $j$ would, while infected, infect county $i$.  
Just as \cref{eq:spatial0} depicts increasingly coarse-grained descriptions of individual-to-individual transmission,
\begin{equation}
\label{eq:spatial1}
R^{(1,1)}\rightarrow R^{(2,1)}\rightarrow R^{(3,1)}
\end{equation}
depicts increasingly coarse-grained descriptions of county-to-county transmission in which the top eigenvalue $R_1$ is preserved. 
Note that, in analogy with individual-to-individual transmission, $R^{(n,1)}_{ij}$ includes all counties in $i$ infected by a county in $j$ from that county's initial infection until its recovery. Since counties typically recover more slowly than single individuals (cf. \cref{fig:self-similarity}), $R^{(n,1)}_{ij}$ describes the system at a larger \emph{time}-scale compared to $R^{(n,0)}_{ij}$.  

A description at the level of states/provinces (third row of \cref{fig:Rmn}) would be appropriate for considering how to stop the spread of disease between states.  A state/province is considered infected if it contains at least one infected county.  In our running example,  $R_2=R^{(3,2)}$ would thus represent the \textit{state-to-state} reproduction number (i.e. the top eigenvalue of $R^{(2,2)}$); in analogy with \cref{eq:spatial0,eq:spatial1},
\begin{equation}
\label{eq:spatial2}
R^{(2,2)}\rightarrow R^{(3,2)},
\end{equation}
where again the time scale associated with $R^{(n,2)}_{ij}$ is typically larger than for $R^{(n,1)}_{ij}$.  
Even higher-scale descriptions (e.g.~considering the nation-to-nation reproduction number) are of course also possible. 

Summarizing \cref{eq:spatial0,eq:spatial1,eq:spatial2}, 
\begin{equation}
\label{eq:grid}
    \begin{array}{cccc}
        R^{(0,0)} \rightarrow & R^{(1,0)}\rightarrow & R^{(2,0)} \rightarrow  & R^{(3,0)} \\
        & \downarrow & \downarrow & \downarrow \\[2pt]
        & R^{(1,1)}\rightarrow & R^{(2,1)}\rightarrow & R^{(3,1)} \\
        & & \downarrow & \downarrow \\[2pt]
        & & R^{(2,2)} \rightarrow & R^{(3,2)}
    \end{array}
\end{equation}
with the horizontal and vertical arrows indicating increasing scale of aggregation and increasing scale of the infectious unit, respectively.

\section{Fractal elimination policies}
\begin{figure*}
    \centering
    \includegraphics[width=\textwidth]{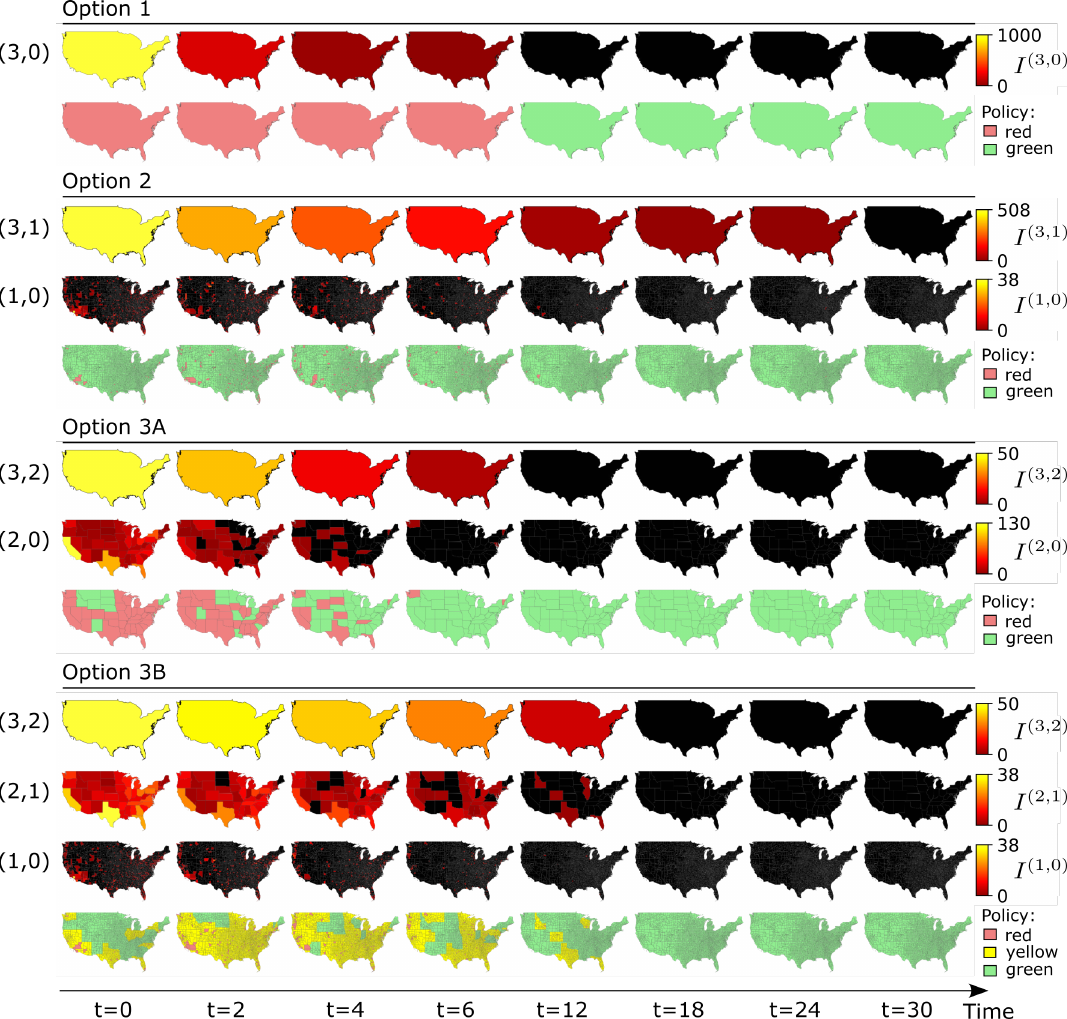}
    \caption{\textbf{Pandemic evolution at different levels of representation.} Map of continental U.S. showing at the $(m,n)$-scale (left labels) the number of $n$-level infections $I^{(m,n)}$ in each $m$-level region and the regional policy at different time steps $t$ (horizontal axis). We used the four different policies described in the text: Option 1 (uniform policy), 2 (county-level policy), 3A (state-level policy), and 3B (county- and state-level policy). In the policy plots, green regions have no restrictions, red regions have local and travel restrictions, and yellow regions have only travel restrictions. The parameters are chosen such that all policies are in the elimination regime. 
    \emph{Simulation parameters:} $R_0=2$, $\kappa=2$, $I_0 = 1000$, $\alpha=0$, $h = 6$, $r_L=5$, $r_T=5$; see Methods. 
    }
    \label{fig:map_evolution}
\end{figure*}

\vspace{-.5em}
Which policies achieve stable elimination?  By way of example, let us consider three options.  Fig.~\ref{fig:map_evolution} shows simulations of these different policies in action as they achieve elimination, using a heterogeneous county-level SIS metapopulation model (our code is available here~\cite{simulations_repo}). The model is constructed using population data~\cite{UScensus_pop_2021}, as well as commuting and flight data~\cite{UScensus_commuting_2011_15,faa_2019}, in order to estimate a matrix of transmission probabilities between counties (see Methods for details).
Multi-scale policies are then implemented by changing transmission probabilities according to which regions are infected.
This approximate model captures important aspects of the heterogeneity, topology and time-evolution of realistic transmission networks in order to test our suggested multi-scale policies.
While we showcase only a few examples, we note that our simulation framework can be used to stochastically simulate any set of population partitions and multi-scale policies; extending the framework to dynamics beyond SIS is also straightforward.

Option 1: $R^{(3,0)} < 1$.  In this case, a single nationwide policy $P^{(3,0)}$ is adopted that reduces the individual-to-individual reproduction number $R_0=R^{(3,0)}$ below $1$ (first group of plots in fig.~\ref{fig:map_evolution}). If this reduction in $R^{(3,0)}$ can be achieved primarily with testing/contact tracing/isolation, as was the case for SARS-CoV-1, it is likely the best option.  However, if a nationwide lockdown is necessary to achieve $R^{(3,0)}<1$, other options for achieving elimination may be far more cost-effective. 

Option 2: $R^{(3,1)} < 1$.   Even if $R_0=R^{(3,0)}>1$, it is still possible to reduce the county-to-county reproductive number $R_1=R^{(3,1)}$ below $1$.    For instance, consider a policy in which restrictions are imposed only in counties experiencing community transmission and are removed once the outbreak is contained.  In this way, most counties have no restrictions most of the time (second group of plots in fig.~\ref{fig:map_evolution}). When a county $j$ is infected, infections will start to increase at first (since $R^{(1,0)}_{jj}>1$), but then measures can be put in place to contain the disease within county $j$ (i.e.~to achieve $R^{(1,0)}_{jj}<1$) and reduce the spread to other counties (i.e.~by reducing the off-diagonal elements $R^{(1,0)}_{ij}$ in column $j$) until county $j$ has recovered (i.e.~no community transmission).   We denote this policy by $P^{(1,0)}$, which will result in an $R^{(1,0)}$ that varies in time depending on which counties are infected.  Because $R^{(1,0)}$ varies depending on infections, its top eigenvalue $R_0=R^{(3,0)}$ is not a good characterization of whether $P^{(1,0)}$ will achieve elimination.  However, just as the time-varying dynamics of viral spread \textit{within} humans can correspond to a constant $R_0$ describing the spread of disease \textit{between} humans, the time-varying spread of the disease \textit{within} counties corresponds to a constant $R_1=R^{(3,1)}$ describing the spread of disease \textit{between} counties, as set by a national policy $P^{(3,1)}$. Writing this policy using the schematic of \cref{eq:grid},
\begin{equation}
\label{eq:r10}
    \begin{array}{cccc}
         & R^{(1,0)} &  & \\
        & \downarrow &  &  \\[2pt]
        & R^{(1,1)} & \xrightarrow{\hspace{5em}} & R^{(3,1)} \\
        & & & 
    \end{array}
\end{equation}
where $R^{(1,1)}_{ij}$ denotes the expected number of times that county $i$ will be infected by county $j$ if $j$ is infected.  This equation illustrates a two-step process in which we first coarse-grain over the counties' time-dependent internal dynamics caused by $P^{(1,0)}$ to derive a constant county-to-county matrix $R^{(1,1)}$ (analogous to coarse-graining the viral dynamics within individuals to the individual-to-individual transmission matrix $R^{(0,0)}$) and then characterize disease spread at the national level by coarse-graining $R^{(1,1)}$ to its top eigenvalue $R_1=R^{(3,1)}$ (see Methods).  

To summarize, we first choose a national policy $P^{(3,1)}$ with a target $R_1=R^{(3,1)}$ and then choose a county-level policy $P^{(1,0)}$ such that coarse-graining leads to our national target of $R^{(3,1)}$.  Note that many possible county-level policies can lead to the same $R^{(3,1)}$, a degree of freedom that can be used to reduce costs.

Option 3: $R^{(3,2)} < 1$.  Depending on the disease, achieving $R_1=R^{(3,1)}<1$ via Option 2 may not be feasible.  One alternative is Option 3A (third group of plots in fig.~\ref{fig:map_evolution}), in which a policy is made at the state rather than county level (e.g. because it may be easier to reduce interstate travel than inter-county travel).  In this case, the national policy $P^{(3,2)}$ specifies a constant state-to-state reproduction number $R_2=R^{(3,2)}<1$.  A policy $P^{(2,0)}$ is then chosen so that the resulting $R^{(2,0)}$ (which will dynamically depend on which states are infected) coarse-grains to $R^{(3,2)}$, analogously to Option 2 except with states in place of counties.  Using the schematic of \cref{eq:grid},
\begin{equation}
    \begin{array}{cccc}
        & ~~~~~~~& ~~~~R^{(2,0)}  &  \\[4pt]
        & ~~~~~~~& ~~\Bigg\downarrow &  \\[16pt]
        & ~~~~~~~& ~~~~~~~~\,R^{(2,2)} \rightarrow & R^{(3,2)}
    \end{array}
\end{equation}
  Both $R_1=R^{(3,1)}$ for Option 2 and $R_2=R^{(3,2)}$ for Option 3A are equivalent  to the region-to-region reproduction number~\cite{Siegenfeld2020}, with the regions being either counties or states, respectively.  

Rather than solely acting at either the state or county level, however, we can also consider a policy Option 3B.  As in Option 3A, a national policy $P^{(3,2)}$ specifies a constant state-to-state reproduction number $R_2=R^{(3,2)}<1$.  Then, we choose a state-level policy $P^{(2,1)}$ that results in the desired value of $R^{(3,2)}$ as follows. When a state $j$ is infected, instead of attempting to get $R^{(2,0)}_{jj}<1$ within the entire state $j$ as in Option 3A, more targeted (and thus potentially lower-cost) county-level actions are taken such that the county-to-county reproduction number within the state $R^{(2,1)}_{jj}<1$ (fourth group of plots in fig.~\ref{fig:map_evolution}). In addition, travel out of state $j$ may also be reduced, corresponding to a reduction of the off-diagonal elements of $R^{(2,1)}_{ij}$ in the $j^{th}$ column, which can reduce spread to other states and thus reduce $R_2$.  We then repeat this process at a smaller scale. That is, just as the state policy $P^{(2,1)}$ is chosen to obtain a particular value for the state-to-state reproduction number $R_2=R^{(3,2)}$ dictated by the national policy $P^{(3,2)}$, county-level policies $P^{(1,0)}$ are chosen for each value of $R^{(2,1)}$ dictated by the state-level policy $P^{(2,1)}$ (the value of $R^{(2,1)}$ depending on which states are infected).  The county-level policy $P^{(1,0)}$ thus dynamically depends on which states are infected and will result in an $R^{(1,0)}$ that dynamically depends on which counties are infected.  In the schematic of \cref{eq:grid}, 
\begin{equation}
\label{eq:r21}
    \begin{array}{ccc}
          R^{(1,0)} & ~ & ~ \\
         \downarrow & ~ & ~ \\[4pt]
          ~R^{(1,1)} & \rightarrow R^{(2,1)}  & ~\\[2pt]
         ~ & \downarrow & ~ \\[4pt]
         ~ & ~~~~R^{(2,2)} & \rightarrow  R^{(3,2)}
    \end{array}
\end{equation}
In other words, some desired value of $R_2=R^{(3,2)}$ is set at the national level, states then determine a policy $P^{(2,1)}$ that reduces the spread of the disease between counties in order to achieve the desired $R^{(3,2)}$, and counties then adopt policies $P^{(1,0)}$ that reduce the spread of disease among individuals in order to achieve the desired (time-varying) values of $R^{(2,1)}$ set at the state level according to the policy $P^{(2,1)}$.  

The above policies (Options 1, 2, 3A, and 3B) sample only a small part of the space of possible multi-scale pandemic responses.  To illustrate just one other possibility, we note that in the strategies considered above, only the infected regions take action.  But just as contact tracing can combat disease spread at the individual level by not only isolating infected individuals but also quarantining their close contacts, a multi-scale contact tracing strategy can also be employed.  Infected regions at various scales can be labeled as red, with their ``close contacts'' (i.e. regions with with they have high rates of travel) labeled as orange.  At the county-level for instance, infected (red) counties would experience both local reduction measures and a restriction of travel/border monitoring out of them, and counties in close contact (orange) would also have travel restricted out of them until testing can establish that they are uninfected.  Similar policies can be implemented at larger scales.

\section{Stable elimination with importations} \label{sec:stable}
We note that elimination, i.e. zero cases, is a fixed point of the system in the absence of external perturbations (e.g.~importations of cases).  The question is whether this fixed point is stable or unstable, i.e. whether or not cases arising from importations will exponentially grow or decline.  The previous section is essentially an attempt at converting highly complex dynamics, in which humans are responding to transmission within regions at various scales, into a larger-scale theory in which transmission can be characterized by a single time-independent matrix.  

This larger-scale theory then gives a very simple criterion for the stability of elimination: elimination is stable if the largest eigenvalue of this matrix is less than $1$~\cite{diekmann1990definition}.  This criterion cannot necessarily be applied at smaller scales, since the smaller-scale next-generation matrices may vary in time depending on which regions are infected---the largest eigenvalue $R_n$ of such matrices may be greater than $1$ at all times, but the disease may still not be able to spread due to the time-varying nature of policy response.  For instance, the policies at the level of individuals ($n=0$) and counties ($n=1$) in \cref{eq:r21} describe a complex pattern of disease spread with policies that change in response to changing conditions, but the spread between states ($n=2$) can be characterized by the single time-independent matrix $R^{(2,2)}$, the largest eigenvalue of which is $R_2=R^{(3,2)}$.  If $R_2>1$, the number of infected states will exponentially grow, while if $R_2<1$, perturbations in the number of infected states will exponentially decay back to zero. 

Formally, let $N$ be the scale of the entire population ($N=3$ in the example above) and $n^*$ be the smallest scale at which, for the policy in question, $R^{(n^*,n^*)}$ (and thus also $R^{(n,n*)}$ for $n>n^*$) has no time-dependence (in the example above, $n^*=2$).  A sufficient condition for stable elimination is then 
\begin{equation}
\label{eq:stability}
    R_{n^*}=R^{(N,n^*)}<1
\end{equation}
 We note that $R^{(N,n^*)}$ will be an overestimate of the spread of disease, since its derivation essentially assumes all entities are susceptible (see Methods for more details); thus, \cref{eq:stability} is a sufficient but not necessary condition for the stability of elimination.  

Suppose a population that has collectively pursued an elimination policy ($R_{n^*}<1$) experiences an importation of the disease from some other population.  Not every importation will result in an outbreak of community transmission; many importations may be contained either by testing and isolation at the border or contact tracing efforts.  But for each importation that does lead to an outbreak, we can bound the expected number of $n^*$-entities that will become infected (and thus have to impose restrictions) by
\begin{equation}
1+R_{n^*}+R_{n^*}^2+R_{n^*}^3+...=\frac{1}{1-R_{n^*}}
\label{eq:import}
\end{equation}
since $R_{n^*}^k$ is an upper bound for the expected number of infected $n^*$-entities in the $k^{th}$ generation.

\begin{figure*}
    \centering
    \includegraphics[width=\textwidth]{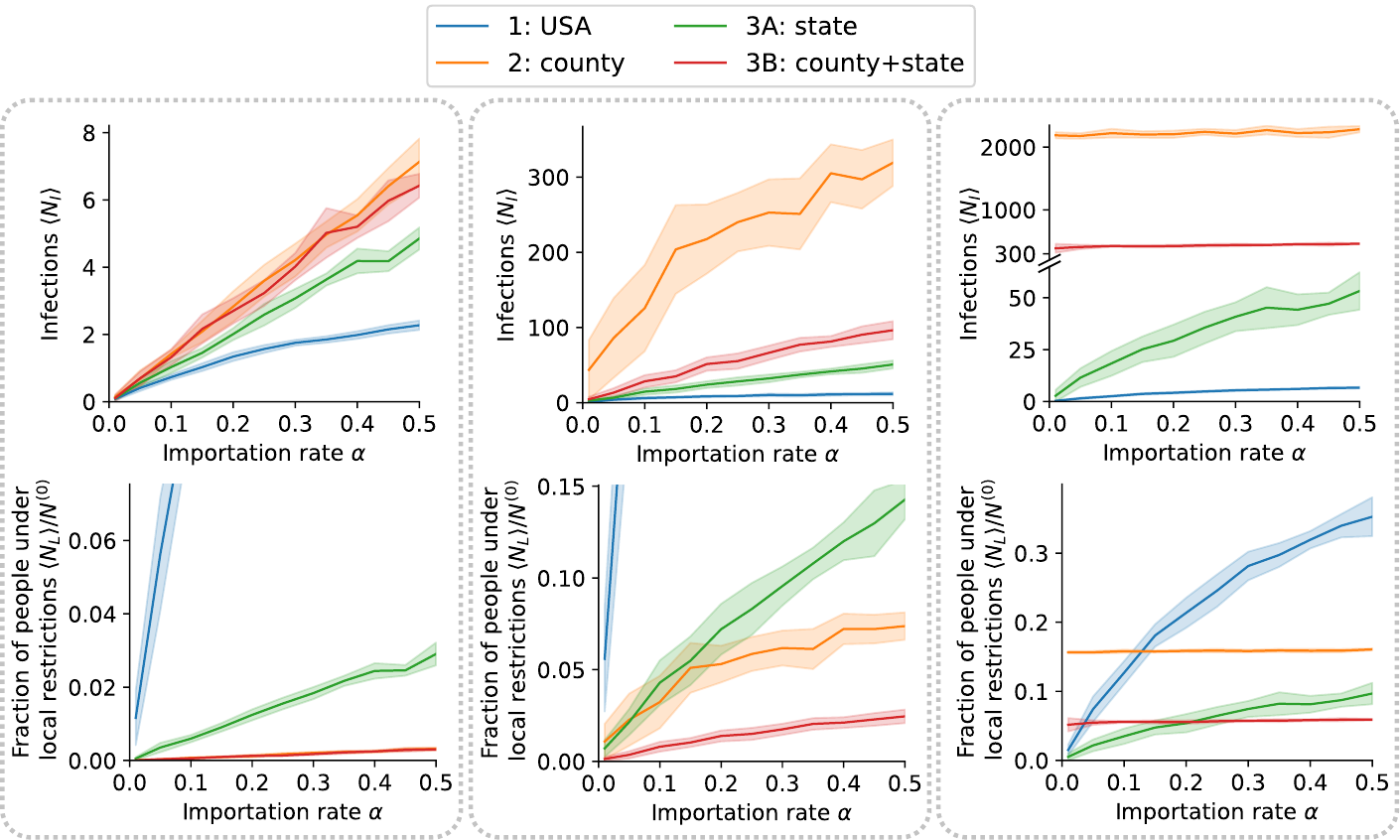}
    \caption{\textbf{Steady-state infections and social restrictions under importation of cases.} Comparison of the average number of cases and average fraction of people experiencing social distancing measures at any given time.  The effect of four different multi-scale policies (Options 1, 2, 3A, 3B) is shown for three different parameter sets.  The system is in the elimination regime when the number of infections and people experiencing restrictions tends to zero as $\alpha \rightarrow 0$.  Options 1 and 3A are in the elimination regime for all three sets of parameters, while Option 2 is in the elimination regime only in the left column and Option 3B only in the left and center columns. 
    \emph{Simulation parameters:} $R_0=2$, $\kappa=2$, $I_0 = 10$ for all. For each column: (left) $h=5$, $r_L=5$, $r_T=5$; (center) $h=10$, $r_L=2.2$, $r_T=10$; (right) $h=20$, $r_L=5$, $r_T=5$. See Methods for details.     }
    \label{fig:cost_effectiveness}
\end{figure*}

Fig.~\ref{fig:cost_effectiveness} displays simulations (using the same data as \cref{fig:map_evolution}) of the number of steady-state infections and the total fraction of people under restrictive measures resulting from the policies of Options 1, 2, 3A, and 3B in the presence of importations.  We show results for three different sets of parameters.  The policies that achieve elimination result in both fewer infections and fewer restrictions than those that do not if the importation rate can be kept sufficiently low.  For the parameters shown, at least one of the multi-scale policies (Options 2, 3A, and 3B) results in a far smaller fraction of the population being under restrictions at any given time than the homogeneous policy (Option 1) while still achieving elimination.

We note that the criterion given in \cref{eq:stability} is robust.  Suppose a policy in which $R_{n^*}<1$ is implemented, but there are unforeseen deviations in implementation (foreseen deviations being already accounted for).  Then, as long as such deviations are not so large as to raise $R_{n^*}$ above $1$,  elimination will remain stable.  

Policies can be designed with a margin for error (e.g. travel restrictions need not be airtight, and local transmission rates need not be zero as long as within the infected region $R_0<1$); thus, individual people violating policy can be accounted for.  It is, however, important that all of the regions containing the population in question cooperate.  Even one uncooperative region can result in $R_{n^*}>1$: for instance, if one state allows the disease to spread rampantly, it will continually infect other states, resulting in an unbounded state-to-state reproduction number.  Thus, in order for elimination to be maintained, any regions unwilling or unable to cooperate must be excluded from the population, and travel from these regions must be restricted.  Then, all of the theory above will apply to the portion of the population that is cooperating, with any cases arising from the uncooperative regions being treated as importations (see \cref{eq:import}).

\section{Minimizing disease burden and economic cost} \label{sec:costs}
 
As discussed above, part of the usefulness of conceptualizing of policy response in this nested way is that the optimal policy does not have to be chosen ``all at once,'' but rather each region can independently choose whichever policy is least costly, subject to the constraint that it results in the appropriate coarse-grained targets. We estimate the cost of a policy as a long-time average under the assumption of external importations (if there are no importations, then any elimination policy incurs only a one-time cost).  Given a policy that achieves stable elimination, as described in the previous section, we can expand the expected value of its cost to first order in the importation rate $\alpha$; in other words, we can write 
\begin{equation}
C(\alpha)=c\alpha+O(\alpha^2)
\end{equation}
for some constant $c$ where $C$ is the expected overall cost of enacting the policy per unit time.  The average number of infected entities at each scale will be approximately proportional to $\alpha$, and each will incur an expected cost from both the disease itself and the measures necessary to contain it. 

For policy functions that result in transmission matrices satisfying \cref{eq:r21}, we can write 
\begin{equation}
\label{eq:c0}
c=c_0[P^{(1,0)},~P^{(2,1)},~ P^{(3,2)}]
\end{equation}
This cost function could be optimized sequentially by defining 
\begin{align}
c_1[P^{(2,1)},P^{(3,2)}]&\equiv \min_{P^{(1,0)}} c_0[P^{(1,0)},P^{(2,1)},P^{(3,2)}] \\
    c_2[P^{(3,2)}]&\equiv \min_{P^{(2,1)}} c_1[P^{(2,1)},P^{(3,2)}] \\
    c^*&\equiv \min_{P^{(3,2)}} c_2[P^{(3,2)}]
\end{align}
In other words, we first optimize the county-level policy $P^{(1,0)}$ constrained to achieving given values of $R^{(2,1)}$, we then optimize the state-level policy $P^{(2,1)}$ constrained by $R^{(3,2)}$ and finally we optimize over the target $R^{(3,2)}$ (i.e. the national-level policy $P^{(3,2)}$).
Policies can also be optimized over supra-national scales (see \cref{fig:self-similarity}), at scales below counties (e.g. cities or neighborhoods) and/or at intermediate scales (e.g. between counties and states or between states and the nation); \cref{eq:c0} is just an illustrative example.

Note that this framework need not assume perfect implementation; expected deviations from the theoretically ideal implementation of any given policy---which may increase with policy complexity---can be built into the cost function $c_0$ such that the expressions for $c_1$, $c_2$, and $c^*$ reflect expected actual costs under real-world conditions.

Optimizing policy over all scales will result in an equal or lower overall cost than restricting policy response to two scales, which in turn will result in an equal or lower overall cost than the optimal single-scale policy; formally, 
\begin{align}
\min_{P^{(1,0)},~P^{(2,1)},...,~ P^{(N,N-1)}} c_0&[P^{(1,0)},~P^{(2,1)},...,~ P^{(N,N-1)}] \nonumber \\
\leq \min_{P^{(n,0)},P^{(N,n)}} 
c_0&[P^{(n,0)},P^{(N,n)}] \nonumber \\
\leq \min_{P^{(N,0)}} c_0&[P^{(N,0)}] \label{eq:opt}
\end{align}
for all $0<n<N$.  This result follows from single-scale policies being special cases of two-scale policies, which are in turn special cases of three-scale policies, and so on (see Methods). 

We note that \cref{eq:opt} will hold regardless of social values.  No matter how one values loss of human health/life as compared with a loss in economic/social activity, the use of multi-scale policies results in equal or (potentially much) lower costs than single- or two-scale policies.  This remains true even when costs associated with the additional complexity of implementing multi-scale policies are accounted for, as such costs can be included in $c_0$.

\section{Discussion} 
Throughout history, disease has often been fought in a geographically heterogeneous manner.  During the COVID-19 pandemic, many countries, such as China and Australia, used a multi-scale approach with geographically targeted measures.  Our work aims to lay out a foundation to understand how such policies work beyond individual-level characterization and thus guide the adoption of more effective and economically efficient policy action against a wide range of pandemic threats, as well as currently endemic diseases.  Ideally, interventions would be as targeted as possible, such as isolating only infected individuals.  Such policies are often insufficient, in which case the next step is usually contact tracing, which still operates at the individual level.  Contact tracing is sometimes sufficient, but when it is not, larger-scale action in which regions are treated analogously to individuals can help.  Travel restrictions (including testing at borders) are analogous to isolating sick individuals, and social distancing measures are analogous to the immune system/medical treatment of the individual; without such measures, the region might never recover and will eventually infect others, since no travel restrictions are perfect.  Region-to-region contact tracing can also be enacted: for instance, if a region is infected, travel can be preemptively restricted from nearby regions while they test their populations.   The framework introduced in this article describes these multi-scale policies in a nested manner, demonstrating that, even with imperfect compliance, they can stop the spread of disease as effectively as homogeneous one-scale measures, but at a potentially much lower cost. 

No epidemic model considers disease spread at the smallest scales---e.g. keeping track of individual atoms and molecules---and indeed, most population-wide models do not even consider the number of viral particles within individuals, instead treating individuals as the fundamental unit.  But just as disease spread is usually modeled as spreading from individual to individual, it can also be modeled as spreading from household to household, neighborhood to neighborhood, city to city, province to province, nation to nation, and continent to continent.  By describing the disease-policy system at multiple scales, we can choose the appropriate scale(s) for policy response, allowing for policy makers to reduce or avoid restrictive measures in as many regions as possible while still ensuring elimination.  

Many works analyzing the tradeoffs between public health and the economy make assumptions that artificially constrain the set of possible policy responses.  In many cases, this tradeoff is illusory, as there exist elimination policies that minimize both health and economic costs.  Fine-grained models often \textit{a priori} foreclose policy options simply by their choice of variables; for instance, even the possibility of heterogeneous policy response is excluded from models that treat populations as well-mixed.  By contrast, even the finest-grain policy description presented in this work, $R^{(1,0)}_{ij}$, still allows great flexibility in how individual counties may go about achieving the levels of internal and external transmission specified by that next-generation matrix.  Contact tracing, precautions for populations at higher risk of infection, neighborhood-by-neighborhood interventions, etc. are all accommodated in addition to any county-wide measures that one may wish to implement. 

The multi-scale reproduction numbers $R_n$ provide robust upper bounds on disease transmission and thus a simple metric for guiding policy action, while allowing for a wide range of complexity at the individual scale.  Thus, the many existing individual-level epidemic models can complement the paradigm of larger-scale modeling that we suggest.  Finer-grained descriptions can be used to determine optimal implementation of coarser-grained ones, with the ultimate choice of coarse-grained parameters in turn depending on their associated fine-grained implementation costs.

\subsection{Acknowledgments}
This material is based upon work supported by the National Science Foundation Graduate Research Fellowship Program under Grant No. 1122374 and by the Hertz Foundation.

\appendix
\section{METHODS}
\section{COARSE-GRAINING}

Coarse-graining is a process where one derives descriptions of higher-scale structures by capturing the most relevant aspects of the dynamics of its smaller-scale components, and ignoring irrelevant details.
Typically, the description obtained in this way is more simple than the original one.

The central quantities in our work are the next-generation matrices $R^{(m,n)}$, where $m\geq n$ gives the level of description.
Recall that $R^{(m,n)}_{ij}$ gives the mean number of $n$-level entities (abbreviated as $n$-entities) in $i$ that an infected $n$-entity in $j$ will infect before it recovers.
We can write
\begin{equation}
    \vec I ^{(m,n)'} = R^{(m,n)}\, \vec{I}^{(m,n)},
\label{eq:lindynamics_mn}
\end{equation}
where $I^{(m,n)}_j$ gives the expected value of the initial number of infected $n$-entities in $m$-entity $j$ and $I^{(m,n)'}_j$ gives the expected next generation of infections resulting from the initial infections in the limit where all $n$-entities are susceptible.  For instance, in the running example in the main text, $I^{(2,1)}_j$ denotes the expected number of infected counties in state $j$, while $I^{(3,1)}$ denotes the expected number of infected counties nationwide.  

In an SIS or SIR model in which we overestimate disease spread by treating all entities as susceptible, we would write
\begin{equation}
\frac{dI^{(m,n)}_i}{dt}=\gamma \sum_j (R^{(m,n)}_{ij}-\delta_{ij})I^{(m,n)}_j
\end{equation}
where $\gamma$ is the recovery rate, but the concept of $R^{(m,n)}$ as a next-generation matrix is more general; the SIS/SIR model is just a particular implementation in which the generation intervals are assumed to be exponentially distributed~\cite{Siegenfeld2022}.  

\subsection{Spatial coarse-graining}
We use the term spatial coarse-graining to describe increasing the scale of the aggregation of infectious units while keeping the scale of the infectious units themselves the same (e.g.~$R^{(0,0)}\rightarrow R^{(2,0)}$ or $R^{(2,1)}\rightarrow R^{(3,1)}$).  We use the term spatial (as opposed to temporal) because the spatial scale that each matrix element represents increases (e.g. from individuals to counties to states to nations) while the temporal scale of transmission does not change. 

We use as an illustrative example the transition from the county-level $R^{(1,0)}$ to the state-level $R^{(2,0)}$.
We start with $\vec{I}^{(1,0)}$, where $I^{(1,0)}_j$ gives the number of infected people in county $j$.
The spatial coarse-graining can be defined by a coarse-graining matrix $g^{(2,1)}$ of size $N_\text{states} \times N_\text{counties}$ which specifies how to group the counties together to form states. Specifically,
\begin{equation}
    g^{(2,1)}_{ij} = \left\{
    \begin{matrix}
    1 &\quad &\text{if county } j \text{ belongs to state } i, \\
    0 &\quad &\text{otherwise.}
    \end{matrix} \right.
\end{equation}
Of course, one could consider more general spatial coarse-graining matrices with weights different from 0 and 1.
Using this we can compute the coarse-grained infection vector $\vec{I}^{(2,0)}$ as
\begin{equation}
    \vec{I}^{(2,0)} = g^{(2,1)} \, \vec{I}^{(1,0)}.
\label{eq:coarsed_I_20}
\end{equation}
$I^{(2,0)}_j$ gives the number of infected people in state $j$.
Clearly, $\vec{I}^{(2,0)}$ contains less information than $\vec{I}^{(1,0)}$.
Notice that there are many vectors $\vec{I}^{(1,0)}$ that correspond to the same coarse-grained vector $\vec{I}^{(2,0)}$.

Next, given the next-generation matrix $R^{(1,0)}$ we would like to derive the next-generation matrix $R^{(2,0)}$ corresponding to a specific coarse-graining $g^{(2,1)}$.
We would like an $R^{(2,0)}$ that fulfills \cref{eq:lindynamics_mn} for $m=2$, $n=0$.
Starting from that equation and using \cref{eq:coarsed_I_20} we get
\begin{equation}
    \vec{I}^{(2,0)'} = R^{(2,0)}\, g^{(2,1)}\, \vec{I}^{(1,0)}.
\label{eq:lindynamics_20}
\end{equation}
On the other hand, multiplying \cref{eq:lindynamics_mn} for $m=1$, $n=0$ from the left by $g^{(2,1)}$ we obtain
\begin{equation}
    \vec{I}^{(2,0)'} = g^{(2,1)}\, R^{(1,0)}\, \vec{I}^{(1,0)}.
\label{eq:lindynamics_10_coarsed}
\end{equation}
We would like for both equations to yield the same $I^{(2,0)'}$, i.e.
\begin{equation}
    R^{(2,0)}\, g^{(2,1)}\, \vec{I}^{(1,0)} \stackrel{!}{=}
    g^{(2,1)}\, R^{(1,0)}\, \vec{I}^{(1,0)}.
\label{eq:coarse_condition}
\end{equation}
which is equivalent to 
\begin{equation}
    R^{(2,0)}\, g^{(2,1)} = g^{(2,1)}\, R^{(1,0)}
\end{equation}
If these equations are satisfied, the state-level dynamics of both representations will be identical.
However, since coarse-graining involves a loss of information, we should not generally expect equivalent state-level dynamics (as reflected by $g^{(2,1)}$ not being invertible).

Thus, we would like to choose which aspects of the dynamics at the county level are most important to keep.
For the purpose of pandemic elimination, the most important information contained in $R^{(1,0)}$ is its maximal eigenvalue~\cite{diekmann1990definition}, 
\begin{equation}
    \lambda \equiv \max \text{eig}(R^{(1,0)}).
\end{equation}
The existence of a unique maximal eigenvalue with an associated eigenvector of all positive entries is guaranteed by the Perron-Frobenius theorem, provided there exists some power $k$ such that all of the entries of ${R^{(1,0)}}^k$ are positive---i.e. that an infection in any county has a non-zero probability of eventually spreading to any other county after sufficiently many generations. 

We would like $R^{(2,0)}$ to preserve the maximal eigenvalue, i.e.~we require that
\begin{equation}
    \lambda = \max \text{eig}(R^{(2,0)}).
\label{eq:max_eig_R20}
\end{equation}
We will denote by $\vec{\psi}_\lambda^{(1,0)}$ the eigenvector of $R^{(1,0)}$ with eigenvalue $\lambda$.

Even though we may not be able to find an $R^{(2,0)}$ that satisfies \cref{eq:coarse_condition}  for all vectors $\vec{I}^{(1,0)}$, we can demand it be true for some vectors $\vec{v}_j$.  Because $g^{(2,1)}$ is full rank, it possesses a right-inverse, which we can, without loss of generality, write as 
\begin{equation}
    W^{(1,2)} \equiv V^{(1,0)}\, (g^{(2,1)}\, V^{(1,0)})^{-1}
\end{equation}
where $V^{(1,0)}$ is an $N_\text{counties} \times N_\text{states}$ matrix that can be written as
\begin{equation}
    V^{(1,0)} = \Big( \vec{v}_1\ \cdots\ \vec{v}_{N_\text{states}} \Big)
\end{equation}
Equation~(\ref{eq:coarse_condition}) applied to these vectors $\vec{v}_j$ hence becomes 
\begin{equation}
    R^{(2,0)}\, g^{(2,1)}\, V^{(1,0)} = g^{(2,1)}\, R^{(1,0)}\, V^{(1,0)}
\end{equation}
which implies
\begin{equation}
    R^{(2,0)} = g^{(2,1)}R^{(1,0)} W^{(1,2)} .
\label{eq:R20_solution}
\end{equation}

We can think of $W^{(1,2)}$ as mapping any given coarse-grained vector $\vec{I}^{(2,0)}$ to a fine-grained vector $\vec{I}^{(1,0)} \equiv W^{(1,2)} \vec{I}^{(2,0)}$, such that $R^{(2,0)}\, \vec{I}^{(2,0)} = g^{(2,1)}\, R^{(1,0)}\, \vec{I}^{(1,0)}$.  Thus, the evolution of each coarse-grained vector exactly corresponds to the evolution of some corresponding fine-grained vector.

In order to choose the column vectors of $V^{(1,0)}$, we impose two conditions (in addition to the condition that $g^{(2,1)}V^{(1,0)}$ is invertible).
First, as stated above, we require the maximal eigenvalue to be preserved by the coarse-graining, \cref{eq:max_eig_R20}.
Secondly, we require that the fine-grained vector $\vec{I}^{(1,0)} \equiv W^{(1,2)} \vec{I}^{(2,0)}$ associated with every possible coarse-grained vector $\vec{I}^{(2,0)}$ contains only non-negative elements, i.e.~it must be a physically realistic vector of expected infections.
These two conditions are fulfilled if we choose $V^{(1,0)}_{ij} = (\vec{v}_j)_i = g^{(2,1)}_{ji} (\vec{\psi}^{(1,0)}_{\lambda})_i$ (recall that $\vec{\psi}^{(1,0)}_{\lambda}$ is the top eigenvector of $R^{(1,0)}$). If we write
\begin{equation}
    \psi^{(1,0)}_{\lambda} = \left( \begin{matrix}
        \vec{w}_1 \\ \vdots \\ \vec{w}_{N_\text{states}}
    \end{matrix} \right),
\end{equation}
where $\vec{w}_j$ contains the entries corresponding to state $j$, then
\begin{equation}
    V^{(1,0)} = \left( \begin{matrix}
        \vec{w}_1 & 0 & \cdots & 0 \\
        0 & \vec{w}_2 & \cdots & 0 \\
        \vdots & \vdots & \ddots & \vdots \\
        0 & 0 & \cdots & \vec{w}_{N_\text{states}}
    \end{matrix} \right).
\label{eq:V_choice_maxeig}
\end{equation}
Note that $\vec{w}_j$ is the projection of $\vec{\psi}^{(1,0)}_{\lambda}$ onto the subspace of counties in state $j$.
(Without loss of generality, this notation assumes that the indices are ordered such that the the counties of any given state have contiguous indices.)

This choice of $V^{(1,0)}$ yields
\begin{equation}
    W^{(1,2)} = \left( \begin{matrix}
        \vec{w}_1/\bar{w}_1 & 0 & \cdots & 0 \\
        0 & \vec{w}_2/\bar{w}_2 & \cdots & 0 \\
        \vdots & \vdots & \ddots & \vdots \\
        0 & 0 & \cdots & \vec{w}_{N_\text{states}}/\bar{w}_{N_\text{states}}
    \end{matrix} \right) ,
\end{equation}
where $\bar{w}_j = \sum_i (\vec{w}_j)_i$ is the sum of infections within state $j$ in the maximal eigenvector.

It is straightforward to show that the above construction leads to an $R^{(2,0)}$ with the same maximal eigenvector as $R^{(1,0)}$.
First, note that the coarse-grained vector $\vec{\psi}^{(2,0)}_\lambda \equiv g^{(2,1)} \vec{\psi}^{(1,0)}_\lambda$ is an eigenvector of $R^{(2,0)}$ with eigenvalue $\lambda$. Second,  by the Perron-Frobenius theorem (which will apply to $R^{(2,0)}$ assuming it applies to $R^{(1,0)}$), $R^{(2,0)}$ will have only one eigenvector with all positive entries, which will correspond to the maximal eigenvalue.  Since $\vec{\psi}^{(2,0)}_\lambda$ has all positive entries, $\lambda$ is the maximal eigenvalue. 

To summarize the spatial coarse-graining procedure and generalize to $R^{(m_1,n)}\rightarrow R^{(m_2,n)}$ for any $m_2>m_1\geq n$ and coarse-graining matrix $g^{(m_2,m_1)}$: (1) find the unique eigenvector $\psi$ corresponding to the largest eigenvalue of $R^{(m_1,n)}$ (existence guaranteed by the Perron-Frobenius theorem), and (2) let $R^{(m_2,n)}=g^{(m_2,m_1)}R^{(m_1,n)}V(g^{(m_2,m_1)}V)^{-1}$ where $V_{ij}=g^{(m_2,m_1)}_{ij}\psi_i$.


\subsection{Temporal coarse-graining}
We use the term temporal coarse-graining to describe the process $R^{(m,n)}\rightarrow R^{(m,m)}$ for any $m>n$ (e.g. $R^{(2,0)}\rightarrow R^{(2,2)}$, $R^{(1,0)}\rightarrow R^{(1,1)}$, $R^{(2,1)}\rightarrow R^{(2,2)}$). Here, the dimensions of the matrix remain the same, as the matrix elements correspond to the same regions, so there is no spatial coarse-graining. However, the scale of the infectious units increases---and with it the temporal scale of transmission.  For instance, the typical time-scale between an individual becoming infected and that individual transmitting the infection is larger than the time-scale over which viral particles multiply. Similarly, the time-scale of individuals infecting each other is smaller than the time-scale between counties infecting each other.

Unlike spatial coarse-graining, which we were able to describe according to a fixed mathematical procedure, temporal coarse-graining will depend on the precise nature and timing of policy response.  Thus, the mathematics to be used would depend on the way in which policy response is approximated.  One example of how individual-to-individual transmission can be temporally coarse-grained to region-to-region transmission for a particular class of policy responses is given in ref.~\cite{Siegenfeld2020}; below, we define the temporal coarse-graining procedure more generally.

Given a transmission matrix $R^{(m,n)}$ (which will dynamically depend on which $m$-entities are infected), we can define the matrix $R^{(m,m)}$ as follows: $R^{(m,m)}_{ij}$ equals the expected number of $n$-entities in $m$-entity $i$ that will be caused by a single infected $n$-entity in $m$-entity $j$ if we ignore all transmission outside of $m$-entity $j$ (i.e. if we consider transmission only from the $j^{th}$ column of $R^{(m,n)}$).

This definition will overestimate the spread of disease, since it treats each infection of an $m$-entity as having equal impact, when in reality, infecting an already-infected $m$-entity might result in significantly less additional spread of disease due to control measures already in place.  For example, infecting individuals in an already-infected region before it recovers will have less of an expected impact than infecting individuals in a susceptible region, in much the same way that spreading viral particles to an already-infected individual will have less of an expected impact than infecting a susceptible individual. 

The fact that this temporal coarse-graining overestimates the spread of disease ensures that the larger-scale reproductive numbers $R_n$ are overestimates and thus that $R_n<1$ is a sufficient condition for the stability of elimination. 

\subsection{MULTI-SCALE POLICIES}
The policies described by $P^{(m,n)}$ vary depending on the state of the system, which in turn varies with time (e.g. regions will adopt different measures depending on whether or not they are infected).  For the sake of simplicity, this dependence was left implicit; here we make it explicit.  

Let $Y_m$ be an object that represents the state of the system as known to policy-makers \textit{at the resolution of $m$-entities}.  A simple version of $Y_m$ would be a binary vector that simply indicates in which $m$-entities community transmission has been detected.  In this case, policy would respond in a binary fashion, as it does in all of the examples given in this text.  More general policies are possible: $Y_m$ could for instance include various types of information on the degree of transmission.

 The running example in the main text assumes that the only levels at which policy-makers wish to consider acting are the city, state/province, and national levels.  However, more generally, one can consider any nested hierarchy with $N$ levels above that of the individual (i.e. there is a single $N$-entity representing the whole system).  Suppose intermediate policy response occurs at a subset of these levels $n_1,n_2,...,n_M$, where $n_0=0$ corresponds to individuals and $n_{M+1}=N$ corresponds to the whole system.  The next-generation matrices $R^{(m,n)}$ as a function of time $t$ can then be written in a self-similar manner: 
\begin{equation}
R^{(n_i,n_{i-1})}(t) = P^{(n_i,n_{i-1})}(Y_{n_i}(t),R^{(n_{i+1},n_i)}(t))
\label{eq:fractal}
\end{equation}
for $1\leq i\leq M$, such that $R^{(n,n-1)}$ coarse-grains to the larger-scale next-generation matrix $R^{(n_{i+1},n_i)}$, i.e. 
\begin{equation}
\label{eq:cnstr}
    P^{(n_i,n_{i-1})}(Y_{n_i}(t),R^{(n_{i+1},n_i)}) \xrightarrow{\text{time}} R^{(n_i,n_i)} \xrightarrow{\text{space}} R^{(n_{i+1},n_i)}
\end{equation}
for some $R^{(n_i,n_i)}$, where $\xrightarrow{\text{space}}$ and $\xrightarrow{\text{time}}$ represent spatial and temporal coarse-graining (described in the previous Methods sections), respectively.  For instance, in a system with scales $n=0,1,2,3,4$ (e.g. individuals, cities, provinces, nations, collection of coordinating nations), a value of $R^{(4,3)}$ can be chosen (which could be constant or could vary depending on the global state of the system $Y_4(t)$), with policies at lower scales defined in terms of higher-scale ones according to \cref{eq:fractal}:
\begin{align}
R^{(4,3)}(t) &= P^{(4,3)}(Y_4(t)) \\
R^{(3,2)}(t) &=P^{(3,2)}(Y_3(t),R^{(4,3)}(t)) \nonumber \\
&=  P^{(3,2)}(Y_3(t),P^{(4,3)}(Y_4(t))) \\ 
R^{(2,1)}(t) &= P^{(2,1)}(Y_2(t),R^{(3,2)}(t)) \nonumber \\ 
&= P^{(2,1)}(Y_2(t),P^{(3,2)}(Y_3(t),P^{(4,3)}(Y_4(t))))  \\ 
R^{(1,0)}(t) &= P^{(1,0)}(Y_1(t),R^{(2,1)}(t)) = \nonumber \\
 P^{(1,0)}&(Y_1(t),P^{(2,1)}(Y_2(t),P^{(3,2)}(Y_3(t),P^{(4,3)}(Y_4(t)))))
\end{align}
such that \cref{eq:cnstr} is satisfied. 

One such policy is as follows: $R^{(4,3)}=R_3$ is the nation-to-nation reproductive number, which, if elimination is desired, should be set by policy makers at some level $R_3<1$.  Nations then choose policies governing province-to-province transmission described by $P^{(3,2)}(Y_3(t),R_3)$ such that a nation-to-nation reproductive number of $R_3$ is achieved.  In a self-similar way,  the city-to-city transmission within and between provinces $P^{(2,1)}(Y_2(t),R^{(3,2)}(t))$ is chosen based on which provinces are infected ($Y_2(t)$) so as to achieve the province-to-province transmission described by $R^{(3,2)}(t)$. And the individual transmission within and between cities $P^{(1,0)}(Y_1(t),R^{(2,1)}(t))$ is chosen based on which cities are infected so as to achieve the city-to-city transmission described by $R^{(2,1)}(t)$.  In this way, pandemic response may be independently optimized at each scale while still ensuring the stability of elimination overall.  Since the information in $Y_2$ and $Y_3$ is contained in $Y_1$, we could have also described $R^{(1,0)}(t)$ solely as a function of $Y_1(t)$, but such a function would be far more complex and would leave implicit the ways in which the policy description could be simplified at larger scales.    

We now prove a more general form of \cref{eq:opt}, namely that for any cost function $c$,
\begin{align}
\min_{...~,~P^{(m,a)},~P^{(b,m)},~...} c&[...~,~P^{(m,a)},~P^{(b,m)},~...] \nonumber \\
\leq \min_{...~,~P^{(b,a)},~...} c&[...~,~P^{(b,a)},~...] \label{eq:opt2}
\end{align}
with $a<m<b$.
To prove \cref{eq:opt2}, we note that, following \cref{eq:fractal,eq:cnstr}, $R^{(m,a)}$ will be a function of $Y_m$ the value of $R^{(b,m)}$ dictated by the policy $P^{(b,m)}$, which will itself depend on $Y_b$ and the value of $R^{(c,b)}$ set by larger-scale policies.   Thus, we can write 
\begin{equation}
R^{(m,a)}(t)=P^{(m,a)}(Y_m(t),P^{(b,m)}(Y_b(t),R^{(c,b)}(t)))
\end{equation}
$R^{(b,a)}$ can be written as
\begin{equation}
R^{(b,a)}(t)=P^{(b,a)}(Y_b(t),R^{(c,b)}(t))
\end{equation}
(In the event $b$ is the largest scale of the system, $R^{(b,a)}\equiv R_a$ and $R^{(b,m)}\equiv R_m$ will just be $1\times1$ matrices and $R^{(c,b)}(t)$ should be removed from the above equations.)

We can then see that any policy described by $P^{(b,a)}$ can also be represented by a $P^{(m,a)}$ that has no dependence on $Y_m(t)$:
\begin{equation}
R^{(m,a)}(t)=P^{(m,a)}(Y_b(t),R^{(c,b)}(t))
\end{equation}
and can thus be spatially coarse-grained 
\begin{equation}
R^{(m,a)}(t)\xrightarrow{\text{space}} R^{(b,a)}(t)
\label{eq:skip}
\end{equation}
(If $R^{(m,a)}$ had depended on $Y_m(t)$, it would not be able to be spatially coarse-grained in such a manner because it would then have a finer-grained dependence on the state of the system then $R^{(b,a)}$, which only depends on the coarser-grained $Y_b(t)$.)
We note that when \cref{eq:skip} can be applied (i.e. if $R^{(m,a)}$ depends only on $Y_b$ and not on $Y_m$), $P^{(b,m)}$ is irrelevant: either $b$ is the largest scale of the system, in which case $R_a=R^{(b,a)}$ is simply the largest eigenvalue of $R^{(m,a)}$, or we can coarse-grain $R^{(m,a)}$ to $R^{(c,b)}$ via 
\begin{equation}
R^{(m,a)}\xrightarrow{\text{space}} R^{(b,a)}\xrightarrow{\text{time}}R^{(b,b)}\xrightarrow{\text{space}}R^{(c,b)}
\end{equation}
rather than
\begin{equation}
 R^{(m,a)}\xrightarrow{\text{time,~space}} R^{(b,m)}\xrightarrow{\text{time}}R^{(b,b)}\xrightarrow{\text{space}}R^{(c,b)}
\end{equation}
Thus, the right-hand side of \cref{eq:opt2} cannot be less than its left-hand side, because the right-hand side minimizes over a subset of the policies minimized over by the left-hand side.

\section{STOCHASTIC SIMULATIONS}
The code used for the simulations of this paper is available in the GitHub repository of ref.~\cite{simulations_repo}.  We use a specific model for the sake of simulation; what follows is one of many possible implementations of the next-generation matrices presented in the main text and previous methods sections.

\section{Summary of simulation parameters}

The parameters for the simulations shown in the main text will be described in the following sections. The most relevant parameters are:
\begin{description}
    \item[$I_0$] Initial seed of infected people, randomly distributed across all USA counties (with probability proportional to the county's population).

    \item[$\alpha$] Importation rate. In each time step, new infections are added randomly across the USA (with probability proportional to the county's population), drawn from a Poisson distribution with mean $\alpha$.

    \item[$R_0$] Individual reproduction number. Each person infects on average $R_0$ people in each time step.

       \item[$\kappa$] Degree of superspreading; see \cref{eq:P_NB}.

    \item[$h$] Threshold of infections (not including infections that have been contained by contact-tracing) above which regions change their policy response. 

    \item[$r_L$] Reduction factor for probability of local transmission (social distancing, masks, etc.).

    \item[$r_T$] Reduction factor for probability of inter-regional transmission (travel).
\end{description}

\section{Multi-scale SIS model}
Recall that we consider a partition of the system into a nested hierarchy of regions of different sizes: individuals ($n=0$), counties ($n=1$), states ($n=2$), USA ($n=3$).
The notation $(m,n)$ means that we are using a representation describing the number of infected $n$-level regions within an $m$-level region.
For example, $(1,0)$ means that we keep track of the number of infected individuals in each county, but not on the specific individuals.
We further define as $N^{(n)}$ the total number of $n$-level regions in the system, and $N^{(m,n)}_i$ as the number of $n$-level regions within the $m$-level region $i$ (note $N^{(n)}\equiv N^{(3,n)}$).
We define the ``parent'' function $\ell: \mathbb{N} \rightarrow \mathbb{N}$ that maps an index $i$ of an $n$-level region to an index $\ell(i)$ of the $(n+1)$-level region that $i$ belongs to. We will often write $\ell_i \equiv \ell(i)$.

In our simulations, we use a discrete-time SIS model, where every individual $k$ at time $t$ has a probability $S_k(t)$ of being susceptible and $I_k(t)$ of being infected.
Since $S_k(t)+I_k(t)=1$ we will only keep $I_k(t)$ in the following.

We assume that an infected person recovers and becomes susceptible again in one discrete time step.
For simplicity, we further assume that an infected person can recover and become re-infected in the same time step.
If person $j$ is infected and person $i$ is susceptible, the probability that $j$ will infect $i$ within one time step is $R^{(0,0)}_{ij}$.
The dynamics are then given by
\begin{align}
I_i(t+1) =&\, 1 - \prod_{j}(1-R^{(0,0)}_{ij}\,I_j(t)).
\label{eq:SISmodel}
\end{align}

We numerically simulate the above SIS model of \cref{eq:SISmodel} stochastically. That is, instead of $I$ denoting a probability, each $n$-level entity is either $I=0$ (susceptible) or $I=1$ (infected).
To move forward in time we use the state of the system at time $t$ to calculate the probabilities of infection at time $t+1$ with \cref{eq:SISmodel} and then use these probabilities to sample the next state from a Bernoulli distribution, where $P_i(1)=I_i(t+1)$.

However, since we consider a large number of individuals (order $10^8$), simulating every single individual is very costly.
Instead, the simulations are done at the (1,0) level, i.e.~we consider the total number of infections in each county.
In this case, to evolve the system we use the state of the system at time $t$ to compute with \cref{eq:SISmodel} the expected number of infections in county $K$ at time $t+1$, i.e.~$\mu_{K} = \sum_{i\in K} I_i(t+1)$.  

To account for the possibility of superspreading we sample the state of the system from a negative binomial distribution~\cite{lloyd-smith_superspreading_2005}, which we parameterize in terms of the mean $\mu$ and a superspreading parameter $\kappa$ as
\begin{equation}
    P_\text{NB}(x;\mu,\kappa) = \begin{pmatrix} x+r-1 \\ x \end{pmatrix} p^r (1-p)^x ,
\label{eq:P_NB}
\end{equation}
where $r = \mu/\kappa$ and $p = 1 / (1+\kappa)$.
Note that the variance of this distribution is $\sigma^2 = \mu (1+ \kappa)$.
This parameterization ensures that adding up negative binomials with the same $\kappa$ but different $\mu_i$ results in another negative binomial with mean $\sum_i \mu_i$ and same $\kappa$.

\section{USA data}

For our simulations we use publicly available USA data on population, commuting, and travel.
Specifically, we use population data from the U.S. Census~\cite{UScensus_pop_2021}, air travel data from the FAA~\cite{faa_2019}, and commuting data from the U.S. Census~\cite{UScensus_commuting_2011_15}.
We also used city-to-county data from ref.~\cite{Simple_maps} and cartographic boundary data from the U.S. Census~\cite{UScensus_geodata_2022}.
The collected data is not sufficient to perform accurate simulations of disease spread in the U.S., but in this paper we are only interested in setting up a model that is qualitatively similar to reality.

The U.S. Census population data~\cite{UScensus_pop_2021} is divided into regions of different sizes, of which we only consider counties and states.
Specifically, there are $N^{(0)}=331893745$ people, $N^{(1)}=3143$ counties, $N^{(2)}=51$ states, and $N^{(3)}=1$ nation.
The different regions are assigned a unique identification number called FIPS.

The U.S. Census residence-to-work commuting data~\cite{UScensus_commuting_2011_15} contains the number of people commuting from a given county A (residence) to county B (work). For simplicity, we remove all entries indicating travel outside the U.S. or commuting to/from Puerto Rico. The data file includes self-flow values, i.e.~commuting within the same county; we ignore these too.
The total number of people commuting outside their county is 39530274.
Even though we further process this data (see ~\cite{simulations_repo}), based on how the data was collected we suspect that we overestimate the amount of inter-county travel in our model.

The air travel data from the FAA~\cite{faa_2019} contains the total number of enplanements for all major airports in the U.S. (we use the 2019 file to avoid changes due to COVID-19).
Each airport is associated to one county.
Enplanements from different airports within the same county are added together.
The total number of enplanements in the U.S. per day is 2540002 passengers.
The absence of international travel in our model means we overestimate national travel.

\section{Transmission matrix from USA data}

To transform the above USA data into a set of transmission probabilities, we assume that the probability $R^{(0,0)}_{ij}$ that person $j$ infects person $i$ only depends on the counties, $\ell_j$ and $\ell_i$, to which they belong, $R^{(0,0)}_{ij} \equiv p_{\ell_i \ell_j}$.
We further assume that, in the absence of interventions, the reproduction number $R_0$ is the same within all counties.

To compute the infection probability between people in different counties we need to transform the commuting and air travel data to probabilities.
The commuting data can be written as a matrix $C$, where $C_{ij}$ gives the number of people commuting daily from county $j$ to county $i$, and we set $C_{ii}=0$.
Note that $C_{ij}$ is not necessarily symmetric.
The FAA air travel data, however, only gives us the total number of enplanements $E_i$ for each county $i$, but it does not tell us where those passengers are going.
Using a simplified model~\cite{Lessler_PLOS2009} we estimate the number of flight passengers traveling from county $j$ to county $i$ per day as
\begin{equation}
	F_{i,j\neq i} = E_j \frac{E_i}{\sum_k E_k}, \quad F_{ii} = 0.
\end{equation}
This means that the probability $F_{ij}/N^{(1,0)}_j$ of a passenger departing from $j$ and arriving in $i$ is independent of the origin $j$ and proportional to how much traffic the destination airport has, i.e.~$E_i$. Since the total number of people $F_{i,j\neq i}$ arriving from $j$ is also proportional to the people leaving $j$, i.e.~$E_j$, this model is more all-to-all than reality: it overestimates the traffic of small airports and underestimates the traffic of big airports~\cite{Lessler_PLOS2009}.

We combine commuting and air travel into a mobility matrix $M_{i,j\neq i}$ which gives the effective number of people from county $j$ that are in county $i$ at any point in time \emph{and} can transmit the disease between counties.
Specifically, we weight the two contribution as
\begin{equation}
	M_{i,j\neq i} = w_F F_{ij}
	+ w_C C_{ij}.
\label{eq:mobility_matrix}
\end{equation}
The prefactors $w_F$ and $w_C$ account for flight transfers, outbound/return flights, the number of days of the time unit, and for the fraction of time that a person from county $j$ spends in county $i$ \emph{and} can get/transmit the disease.
Typically, we use $w_F\approx 1$ and $w_C=1/4$.

For practical purposes, it is useful to define the diagonal of the mobility matrix, which is the effective number of residents that stay in their region as
\begin{equation}
	M_{ii} = N^{(1,0)}_{i} - \sum_{j \neq i} M_{ji}.
\end{equation}
This definition ensures $N^{(1,0)}_{i} = \sum_i M_{ij}$. However, note that if $M_{ij}$ is not symmetric, then $\tilde{N}^{(1,0)}_{i} \equiv \sum_{j} M_{ij} \neq N^{(1,0)}_{i}$, where $\tilde{N}^{(1,0)}_{i}$ is the effective number of people present in county $i$ at any moment in time (in our model).

We will assume a well-mixed model where the number of people in each county and the number of commuters/air-travelers is constant at all times.
We transform the mobility matrix into transmission probabilities as
\begin{equation}
	R^{(0,0)}_{ij} = \sum_K P(i\text{ in }K)\, P(j\text{ in }K)\, P(\text{local infection in }K),
\end{equation}
where the sum is over all counties $K$.
This expression assumes that person $j$ can infect person $i$ in any county $K$ as long as they are both present in $K$ for some time.
Specifically, the probability for local infection is
\begin{equation}
	P(\text{local infection in }K) = R_0 / \tilde{N}^{(1,0)}_{K}.
\end{equation}
Note that we use here $\tilde{N}^{(1,0)}_{K}$ instead of $N^{(1,0)}_{K}$.
The probability that a person is in a given county $K$ is
\begin{equation}
	P(i\text{ in }K) = M_{K\ell_i} / N^{(1,0)}_{\ell_i}.
\end{equation}
Notice that the diagonal element of $M$ ensures that $\sum_{K} P(i\text{ in }K) = 1$.
Furthermore, with the above definitions we ensure that $\sum_{i} R^{(0,0)}_{ij} = R_0$, i.e.~the reproduction number is independent from the mobility matrix and the county; what changes is who gets infected, not how many.

\section{One-scale policies}

We consider one-scale policies (Option 1 in main text) where the whole nation (USA) is either red ($R$) or green ($G$). When the number of infections rises above a given threshold $h$, the nation switches to red and we reduce $R_0$ by a factor $r_L$. When cases fall back to zero again, the nation switches to green and we return $R_0$ to its original value.

\vspace{-1em}
\section{Two-scale policies}

We consider two-scale policies (Option 2 and 3A in main text) where regions of any level $n$ are categorized as either red ($R$) or green ($G$), and two types of restrictions are applied: local restrictions within a region (lockdowns, masks, etc.), and travel restrictions between regions (travel reductions, quarantine, testing requirements, etc.).
The rules are as follows: a red region ($R$) has local restrictions, whereas a green region ($G$) does not; if two regions are both green ($GG$) then there are no travel restrictions, but if either of them is red ($GR$, $RG$, or $RR$) travel restrictions apply.

A green region switches to red when the number of local infections rises above a threshold $h$. A red region switches back to green when the number of local infections reaches exactly zero.
When a green region $K$ switches to red we decrease the probability of local transmission by a factor $r_L$, i.e.~$R^{(0,0)}_{ij} \rightarrow R^{(0,0)}_{ij}/r_L$ for $i,j\in K$.
Given a pair of green regions $K,L$, if one or both of them switch to red we reduce the inter-county transmission probability by a factor $r_T$, i.e.~$R^{(0,0)}_{ij} \rightarrow R^{(0,0)}_{ij}/r_T$ for $i\in K$, $j\in L$ or vice versa.
When the inverse transitions happen, i.e.~$R\rightarrow G$ or $GR, RG, RR \rightarrow GG$, we restore the transmission probabilities to the original value.

Note that in simulations with a \emph{two-scale} policy the policy of the higher scale corresponds to the whole nation and is implicit in the value chosen for $R_0$.

\vspace{-.5em}
\section{Three-scale policies}
\vspace{-.5em}
We consider three-scale policies (Option 3B in main text) at levels $n_1$ and $n_2>n_1$ and categorize regions of either level as red ($R_{1,2}$) or green ($G_{1,2}$).
A region of the lower level $n_1$ becomes red ($G_1\rightarrow R_1$) if the number of infections in that region rises above a threshold $h$, and it goes back to green ($R_1\rightarrow G_1$) if the number of infections in the region reaches exactly zero.
This is exactly the same as for the two-scale policy.
A region of the higher level $n_2$ instead becomes red ($G_2 \rightarrow R_2$) if one of its constituent sub-regions of level $n_1$ becomes red.
Conversely, the $n_2$-level region becomes green again ($R_2 \rightarrow G_2$) if all its constituent $n_1$-level sub-regions become green.

Similar to the two-scale policy, in the three-scale policy local restrictions lead to a reduction of the corresponding transmission probabilities by a factor of $r_L$, and travel restrictions by a factor of $r_T$.
In our policy, a region of the lower level $n_1$ applies local restrictions if it is red, and lifts them when it becomes green.
Given two $n_1$-level regions, if either or both are red then travel restrictions are applied.
So far, it is all the same as for the two-scale policy.
The difference now is that if an $n_2$-level region is red, then we apply travel restrictions to all its constituent $n_1$-level sub-regions, regardless of the status of those sub-regions.
This means that travel restrictions apply between two $n_1$-level regions if either or both of them is red ($R_1$), or if either or both of the $n_2$-level regions to which they belong is red ($R_2$).
Travel is unrestricted when both $n_1$-level regions are green and their $n_2$-level parent regions are green too.
Notice that local restrictions at the $n_1$-level still only depend on the state of the $n_1$-level region ($G_1, R_1$).

As before, in simulations with a \emph{three-scale} policy the policy of the higher scale corresponds to the whole nation and is implicit in the value chosen for $R_0$.

\bibliography{bib}

\end{document}